\documentclass[review,12pt]{elsarticle}
 
\usepackage{amssymb}
\usepackage{natbib}
\usepackage{nomencl}
\usepackage{multirow}
\usepackage{makeidx}
\usepackage{enumerate}
\usepackage{rotating}
\usepackage{threeparttable}
\usepackage{color}
\usepackage{array}
\usepackage{amsmath} 
\usepackage[table]{xcolor} 
\usepackage{booktabs} 
\usepackage{amssymb}
\usepackage{nomencl}
\usepackage{amsmath} 
\usepackage{graphicx}
\usepackage{multirow}
\usepackage{anysize}

\journal{Renewable \& Sustainable Energy Reviews}

\begin{document}

\begin{frontmatter}



\title{Decision making under uncertainty in energy systems: state of the art}
\author[focal]{Alireza Soroudi\corref{cor1}}
\ead{University of Tehran, Tehran, Iran, Tel :(Office) +98(21) 66121431
Fax :	+98(21) 66124160, North kargar Street, Tehran, Iran}
\author[focal2]{Turaj Amraee}



\cortext[cor1]{Corresponding author}
\address[focal]{Faculty of New Sciences and Technologies, University of Tehran, Tehran, Iran. Email: soroudi@ut.ac.ir.}
\address[focal2]{Department of Electrical Engineering, K.N. Toosi University of Technology, Tehran,Iran Email: amraee@kntu.ac.ir.}









\begin{abstract}
The energy system studies include a wide range of issues from short term (e.g. real-time, hourly, daily and weekly operating decisions) to long term horizons (e.g. planning or policy making). The decision making chain is fed by input parameters which are usually subject to uncertainties. The art of dealing with uncertainties has been developed in various directions and has recently become a focal point of interest. In this paper, a new standard classification of uncertainty modeling techniques for decision making process is proposed. These methods are introduced and compared along with demonstrating their strengths and weaknesses. The promising lines of future researches are explored in the shadow of a comprehensive overview of the past and present applications. The possibility of using the novel concept of Z-numbers is introduced for the first time.
\end{abstract}

\begin{keyword}
Fuzzy arithmetic\sep info-gap decision theory\sep probabilistic modeling\sep robust optimization\sep interval based analysis \sep Z-number. 
\end{keyword}

\end{frontmatter}

\section{Introduction}
\label{sec:introduction}
The uncertainty handling has been one of the main concerns of the decision makers (including governors, engineers, managers, and scientists) for many years \cite{attoh2005applied}. Most of the decisions to be made by energy sector decision makers are subject to a significant level of data uncertainty \cite{conejou}. 
The uncertain parameters in power system studies can be generally classified into two different categories including (see Fig.\ref{fig:flow}):

\begin{itemize}
	\item {Technical parameters: these parameters are generally categorized in two main classes, namely: topological parameters and operational parameters. The topological parameters are those related to network topologies like failure or forced outage of lines, generators or metering devices and etc. The operational parameters are tied with operating decisions like demand or generation values in power systems.} 
	\item {Economical parameters: the parameters which affect the economical indices fall in this category. Microeconomics investigates the decisions of smaller business sectors like aggregators, domestic or industrial consumers while macroeconomics focuses on entire power system industry. For example, uncertainty in fuel supply, costs of production, business taxes, labor are raw materials are analyzed in microeconomics. On the other hand, the issues like regulation or deregulation, environmental policies, economic growth, unemployment rates, gross domestic product (GDP) and interest rates are analyzed in macroeconomics. All of these parameters are subject to uncertainties and should be correctly addressed in economical studies.}
\end{itemize}

There are various uncertainty handling tools developed for dealing with the aforementioned uncertain parameters as depicted in Fig.	\ref{fig:methods}. The main difference between these methods is in line with the technique used for describing the uncertainty of input parameters. The similarity of them is that all of them try to quantify the effect of input parameters on model's outputs. These models are described as follows:

\begin{itemize}
	\item {Probabilistic approach: one of the earliest works in stochastic programming was done by Dantzig in 1955 \cite{dantzig2011linear}. It is assumed that the input parameters of the model are random variables with a known probability density function (PDF).}
	\item {Possibilistic approach: the fuzzy arithmetic was introduced by Lotfi A. Zadeh in 1965 \cite{zadeh1965fuzzy}. The input parameters of the model are described using the membership function (MF) of input parameters.}
\item {Hybrid possibilistic-probabilistic approaches: both random and possibilistic parameters are present in the model.}
\item {Information gap decision theory:  it was first proposed by Yakov Ben-Haim \cite{IGDT} in 1980. In this method, no PDF or membership function is available for input parameters. It is based on the difference between what is known and what is vital to be known by quantification of severe lack of information in decision making process.}
\item {Robust optimization: it was first proposed by Soyster \cite{Soyster} in 1973. The uncertainty sets are used for describing the uncertainty of input parameters. Using this technique, the obtained decisions remain optimal for the worst-case realization of the uncertain parameter within a given set.}
\item {Interval analysis: it was introduced by Ramon E. Moore in 1966 \cite{moore1966interval}. It is assumed that the uncertain parameters are taking value from a known interval. It is somehow similar to the probabilistic modeling with a uniform PDF. This method finds the bounds of output variables.}
\end{itemize}
\indent This paper is to provide a summary of recent techniques used for uncertainty modeling in power system applications. \textcolor{black}{It offers a vision obtained from a relatively large number of previous works. This review serves as a road map to those interested in uncertainty modeling tools in power system studies to find the less explored research areas by standing on the shoulders of giants.} \\
\indent The rest of this paper is set out as follows: section \ref{sec:prum} presents the Probabilistic approach, 
the possibilistic methodology is introduced in section \ref{sec:psum}, 
the hybrid possibilistic-probabilistic approach is described in section \ref{sec:pmppa}, 
the info-gap decision theory is explained in section \ref{sec:igdt}, the robust optimization technique is described in section \ref{sec:Ro}. Section \ref{sec:IO} presents the interval analysis approach. Section \ref{sec:trends} describes the promising lines of future researches. Finally, section \ref{sec:conclu} summarizes the findings of this work.

\section{Probabilistic approach}
\label{sec:prum}
In the probabilistic approach, a multivariate function, namely $y$, $y=f(Z)$ is available. $Z$ is a vector of the form $Z=[z_1,...,z_m]$, in which $z_1$ to $z_m$ are random parameters with known PDFs while the PDF of $Z$ is tried to be identified. Three probabilistic uncertainty modeling techniques are described as follows:
 
\subsection{Monte Carlo Simulation (MCS)}
\label{ssec:mcs}
The Monte Carlo simulation is carried out in following steps \cite{mcs}. It is assumed that the $z_i$ are uncertain parameters. A sample, $z^{e}_i$, is generated for each input parameter $z_i$, using its PDF. 
The value of $y^e$ as the outcome variable, is calculated using $y^{e}=f(Z^{e})$ where $Z^e=[z^e_1,...,z^e_m]$. The procedure is repeated for a number of iterations, $N_{MC}$. Finally, the outcomes are analyzed using statistic criteria, histograms, confidence intervals and etc. There are some methods for reducing the computational burden of MCS like Latin Hypercube Sampling (LHS) \cite{4808223}. 
\subsection{Point estimate method}
\label{ssec:tpem}
The point estimate method (PEM) acts based on the concept of moments of uncertain input parameters. In a problem with $n$ uncertain parameters, the major steps are as follows \cite{Hong1998261}:
\begin{enumerate}[Step.1]
	\item Set $E(Y)=0 , E(Y^2)=0$ and $k=1$.
	\item {Determine the locations and probabilities of concentrations,  $\epsilon_{k,i}$ and $P_{k,i}$, respectively as follows: 
	\begin{alignat}{2}
&	\epsilon_{k,i}=\frac{1}{2}\frac{M_3(z_k)}{\sigma^3_{z_k}} +(-1)^{i+1} \sqrt{n+\frac{1}{2}(\frac{M_3(z_k)}{\sigma^3_{z_k}})^2}\\
&P_{k,i}=(-1)^i\frac{\epsilon_{k,3-i}}{2n\sqrt{n+\frac{1}{2}(\frac{M_3(z_k)}{\sigma^3_{z_k}})^2}}
	\end{alignat}
where $M_3(z_k)$ is the third moment of parameter $z_k$. 	
	} 
	\item {Determine the concentration points $z_{k,i}$, as given below. 
	\begin{eqnarray}
	z_{k,i}=\mu_{z_k}+	\epsilon_{k,i} \times \sigma_{z_i}, i=1,2
	\end{eqnarray} 
	where, $\mu_{z_k}$ and $\sigma_{z_k}$ are mean and standard deviation of $z_k$, respectively.}
	\item {Calculate the $f$ for both $z_{k,i}$, as:
	\begin{eqnarray}
	Z=[z_1,z_2,...,z_{k,i},...,z_n], i=1,2
	\end{eqnarray}}
	\item {Calculate $E(Y)$ and  $E(Y^2)$ using: 
	\begin{alignat}{2}
	&E(Y)=E(Y)+\sum_{i=1}^2 P_{k,i}f(z_1,z_2,...,z_{k,i},...,z_n)\\
	&E(Y^2)=E(Y^2)+ \sum_{i=1}^2 P_{k,i}f^2(z_1,z_2,...,z_{k,i},...,z_n) \nonumber
	\end{alignat}}	
	\item {$k=k+1$ if $k<n$ then go to Step. 2; otherwise continue.}
		\item {Calculate the mean and the standard deviation as: 
		\begin{alignat}{2}
	&\mu_Y= E(Y)\\
	&\sigma_Y=\sqrt{E(Y^2)-E^2(Y)} \nonumber
	\end{alignat}}
\end{enumerate}

\subsection{Scenario based decision making}
\label{ssec:scenred}
A scenario is defined as a probable realization of an uncertain set of parameters. A list of scenarios are generated using the PDF of each uncertain parameter, $Z_s$. The expected value of output variable, $y$, is calculated as follows:
\begin{eqnarray}
y=\sum_{s \in \Omega_J} \pi_s \times f(Z_s)
\end{eqnarray}
where $\sum_{s \in \Omega_J} \pi_s = 1$ and $\pi_s$ is the probability of $s^{th}$ scenario. \\
If the number of scenarios are large then it is needed to obtain a small set of scenarios representing the original one. The purpose is to select a small set, $\Omega_S$, with the cardinality of $N_{\Omega_S}$, from the original set, $\Omega_J$ \cite{4801580}. A reasonable trade off must be respected between the loss of the information and decreasing the computational burden \cite{conejou}. The scenario reduction technique is carried out via the following steps \cite{5473192,6128764}: 
\begin{enumerate}[step. 1]
	\item Construct the probability distance matrix containing the distance between each pair of scenarios $c(s,\acute{s})$
	\item {Select the fist scenario $s_1$ as follows: 
	    \begin{alignat}{2}
    \label{eq:w1}
	        &  s_1=arg\left\{  \min_{s'\in \Omega_J} \sum_{s\in \Omega_J} \pi_s c(s,s')     \right\}\\
	       &   \Omega_S=\left\{s_1 \right\} , \Omega_J=\Omega_J - \Omega_S \nonumber
	     \end{alignat}  }
		\item {Select the next scenario for $\Omega_S$ set, as follows: 
		   \begin{alignat}{2}
    \label{eq:wn}
	          &s_n=arg\left\{  \min_{s'\in \Omega_J} \sum_{s\in \Omega_J-\left\{s'\right\}}
	          \pi_s  \min_{s'' \in \Omega_S \cup\left\{s \right\}} c(s,s'')     \right\}  \\
	          &\Omega_S=\Omega_S \cup \left\{s_n \right\} , \Omega_J=\Omega_J - \Omega_S \nonumber
     \end{alignat}  }
\item If the cardinality of $\Omega_S$ is sufficient then go to step 2; else continue.
\item Add the probability of each non-selected scenario to its closest scenario in the selected set, End.
\end{enumerate}
More details can be found in \cite{conejou}.

\section{Possibilistic approach}
\label{sec:psum}
Since the introduction of fuzzy set theory this technique has been used in many power system fields \cite{466473}. Suppose $y=f(x_1,\ldots, x_n)$ is in hand and $X$ vector contains the uncertain input parameters described using their associated membership functions. Various membership functions can be used to formulate the degree of membership of a specific uncertain parameter depending on the expert's opinion. Regardless of the membership function's shape the questions is ``how to determine the MF of $y$ if MFs of $X$ are known?''.  
The $\alpha$-cut method can provide an answer to this question \cite{fuzzycontrol}. For a given fuzzy set $\tilde{A}$ in $U$, the crisp set $A^\alpha$ contains all individuals of $U$ with membership value,  $\tilde{A}$, greater than or equal to $\alpha$, as calculated in (\ref{eq:alpha}).
\begin{eqnarray}
\label{eq:alpha}
&&A^\alpha=\left\{x\in U \mbox{ } |\mbox{ } \mu_A(x)\geq \alpha \right\} \\ 
&&A^\alpha=(\underline{A}^\alpha,\bar{A}^\alpha) \nonumber
\end{eqnarray}
The $\alpha$-cut of each uncertain parameter, $x^\alpha_i$, is determined using (\ref{eq:alpha}), then the $\alpha$-cut of y, $y^\alpha$, is calculated as follows:
\begin{alignat}{2}
\label{eq:output}
&y^\alpha=(\underline{y}^\alpha,\bar{y}^\alpha) \\
&y^\alpha=(\min_{X^\alpha} f(X^\alpha),\max_{X^\alpha} f(X^\alpha)) \\
&  X^\alpha=(\underline{X}^\alpha,\bar{X}^\alpha)
\end{alignat}
In each $\alpha$-cut, the upper bound of $y^\alpha$, $\bar{y}^\alpha$, and the lower bound of $y^\alpha$, $\underline{y}^\alpha$, are maximized and minimized respectively. 
The final step is defuzzification. The process of translating a fuzzy number to a crisp one is called defuzzification \cite{fuzzycontrol}.
Many defuzzification techniques are available such as maximum defuzzification technique, the centroid method \cite{deffuz}, weighted average defuzzification technique and etc. 



\section{Hybrid possibilistic-probabilistic approach}
\label{sec:pmppa}
Sometimes, the decision maker is faced with a multivariate objective function, $y=f(X,Z)$, where both possibilistic uncertain parameters ($X$) and probabilistic uncertain ones ($Z$) exist. To deal with such cases some methods are developed which are decsribed next. 

\subsection{Possibilistic-Monte Carlo approach}
\label{ssec:mpc}
The following steps describe the mixed possibilistic-Monte Carlo approach \cite{Soroudi2011794}:
\begin{itemize}
\item Step.1 : For each $z_i \in$ Z, generate a value using its PDF, $z^e_i$
\item Step.2 : Calculate $(\bar{y}^{\alpha})^e$ and $(\underline{y}^{\alpha})^e$ as follows: 
\begin{alignat}{2}
&(\underline{y}^{\alpha})^e=\min f(Z^e,X^\alpha)\\
&(\bar{y}^{\alpha})^e=\max f(Z^e,X^\alpha)\\
&X^\alpha=(\underline{X}^\alpha,\bar{X}^\alpha) 
\end{alignat}
\end{itemize} 
These steps are repeated to obtain the statistical data of the parameters of the output's MF such as PDF or expected values. 
\subsection{Possibilistic-scenario based approach}
\label{ssec:mpcba}
The following steps describe this approach \cite{6142135}:
\begin{itemize}
\item Step.1 : Generate the scenario set describing the behavior of $Z$, ${\Omega_J}$
\item Step.2 : Reduce the original scenario set to a small set, ${\Omega_s}$
\item Step.3 : Calculate $(\overline{y}^{\alpha})$ and $(\underline{y}^{\alpha})$ as follows: 
\begin{alignat}{2}
&\underline{y}^{\alpha}=\min \sum_{s \in\Omega_s} \pi_s \times f(Z_s,X^\alpha)\\
&\overline{y}^{\alpha}=\max \sum_{s \in\Omega_s} \pi_s \times f(Z_s,X^\alpha)  \\
&X^\alpha=(\underline{X}^\alpha,\overline{X}^\alpha)
\end{alignat}
\item Step.4 : Deffuzzify the $y$.
\end{itemize}


\section{Information Gap Decision Theory}
\label{sec:igdt}
The Information Gap Decision Theory (IGDT) is a method to describe the uncertainties which can not be described using PDF of MF due to the lack of sufficient information. It is used to make robust decisions against sever uncertainty of input parameters. In IGDT, the robustness is defined as the immunity of satisfaction of a predefined constraint \cite{IGDT}. The constraint satisfaction is defined based on the requirement of the decision maker. It may be defined as the maximum load which a bridge can tolerate or the maximum risk that the decision maker can accept or even the minimum revenue a decision maker is willing to achieve. Here, the main goal is not only optimizing the objective function \cite{6338333}. 
Instead, the algorithm tries to find the best possible solution which maximum robustness against the probable forcasting errors.\\ \indent The constraint satisfaction is defined as not violating a predefined critical limit, $\zeta$, for a given cost function, $f(x,\bar{d})$, as follows:
\begin{alignat}{2} 
&f(x,\bar{d}) \leq \zeta \\    
& H(x,\bar{d})=0 \\ 
& G(x,\bar{d})\geq 0  
\end{alignat}
where, $x$ is the input parameter and $\bar{d}$ is the vector of decision variables. $H$ and $G$ are the equality and inequality constraints, respectively.\\
\indent The uncertainty of parameters in IGDT method, is usually defined as the envelope bound model \cite{6298065}, as follows:
\begin{alignat}{2}
\label{eq:uncertianty} 
& x \in U(\alpha,\tilde{x}) \\
& U(\alpha,\tilde{x})=\left|\frac{x-\tilde{x}}{\tilde{x}} \right|\leq \alpha  \nonumber
\end{alignat}
where, $\alpha$ is the uncertainty level of parameter $x$, $\tilde{x}$ is the forcasted value of x and $U(\alpha,\tilde{x})$ is the set of all values of $x$ whose deviation from $\tilde{x}$ will never be more than $\alpha \tilde{x}$. The decision maker does not know the values of $x$ and $\alpha$. \\
\indent The robustness of a decision $\bar{d}$ based on the requirement $\zeta$, $\hat{\alpha}(\bar{d},\zeta)$, is defined as the maximum value of $\alpha$ at which the decision maker is sure that the required constraints are always satisfied \cite{IGDT}, as follows:
\begin{alignat}{2}
&\hat{\alpha}(\bar{d},\zeta)=\max \alpha \\
&S.t: \mbox{Constraints} \nonumber
\end{alignat}
The decision making policy is defined as finding the decision variables, $\bar{d}$, which maximizes the robustness, as :
\begin{alignat}{2} 
\label{eq:robust}
&\max_{\bar{d}} {\hat{\alpha}(\bar{d},\zeta)} \\
&\forall x \in U(\alpha,\tilde{x}) \\
& f(x,\bar{d}) \leq \zeta  \\   
&H(x,\bar{d})=0\\
&G(x,\bar{d})\geq 0  
\end{alignat}

%

\section{Robust optimization}
\label{sec:Ro}
The concept of robust optimization (RO) was first introduced by Soyster \cite{Soyster}. It's a new approach for solving optimization problems affected by uncertainty specially in case of lack of full information on the nature of uncertainty \cite{bental}. The concept of robust optimization is described as follows: 
consider a function like $z=f(x,y)$ which is linear in $x$ and non-linear in $y$. The values of $x$ are subject to uncertainty while the values of $y$  are known. In robust optimization, it is assumed that no specified PDF is in hand for describing the uncertain parameter $x$. The uncertainty of $x$ is modeled with an uncertainty set $x \in U(x)$, where $U(x)$ is a set that parameter $x$ can take value from it. 
The maximization of $z=f(x,y)$ can be formulated via (\ref{eq:roz}) to (\ref{eq:roz2}).
\begin{alignat}{2}
\label{eq:roz}
&\max_{y} z= f(x,z)\\
\label{eq:roz2}
& x \in U(x) 
\end{alignat}
Since the value of $z$ is linear with respect to $x$, it can be reformulated as follows:
\begin{alignat}{2}
\label{eq:roz33}
&\max_{y} z \\
&z \leq f(\tilde{x},z) \\
&f(\tilde{x},y)=A(y)*\tilde{x}+g(y) \\
& \tilde{x} \in U(x)=\left\{x| \left|x-\bar{x} \right| \leq \hat{x}  \right\}  
\end{alignat}
where $\tilde{x},\bar{x},\hat{x}$ are the uncertain value, predicted value and maximum possible deviation of variable $x$ from $\hat{x}$, respectively.
The robust optimization seeks a solution which not only maximizes the objective function $z$ but also insures the decision maker that if there exist some prediction error about the values of $x$, the $z$ remains optimum with high probability \cite{priceofro}. To do this, a \textit{robust counterpart} version of the problem is constructed and solved. 
The robust counterpart of (\ref{eq:roz2}) is defined as follows:
\begin{alignat}{2}
\label{eq:roz4}
&\max_{y} z \\
&z \leq f(x,z) \\ \label{eq:ro66}
&f(x,y)=A(y)*\bar{x}+g(y)-\max_{w_i} \sum_i a_i(y)*\hat{x}_i*w_i \\ \label{eq:ro7}
&\sum_{i} w_i \leq \Gamma\\
& 0 \leq {w_i} \leq 1  
\end{alignat}
Based on (\ref{eq:roz4}), two nested optimization problems are to be solved. The equations  (\ref{eq:ro66}) to (\ref{eq:ro7}) are linear with respect to $w_i$ and has a dual form as follows:
\begin{alignat}{2}
\label{eq:roz6}
&\min_{\xi_i,\beta} [\Gamma \beta + \sum_i \xi_i] \\
& \beta +{\xi_i} \geq a_i(y)*\hat{x}_i \nonumber
\end{alignat}
Inserting the (\ref{eq:roz6}) into (\ref{eq:roz4}) gives :

\begin{alignat}{2}
\label{eq:roz7}
&\max_{y,\xi_i,\beta} z \\
&z \leq f(x,z) \\
&f(x,y)=A(y)*\bar{x}+g(y)- \Gamma \beta - \sum_i \xi_i \\
& \beta +{\xi_i} \geq A(y_i)*\hat{x}_i 
\end{alignat}
There are some softwares developed for solving the robust optimization based problems \cite{rosoft}.
%

\section{Interval analysis}
\label{sec:IO}
In this method, the range of values for each uncertain input parameter is defined and it can be represented by an interval. Suppose a multivariate function of the form $f=(x_1,...,x_n)$ and $lb_i \leq x_i \leq ub_i$ where $lb_i,ub_i$ are the lower and upper bounds of uncertain parameter $x_i$. The goal is finding the lower and upper bounds of objective function $f$. There are some softwares developed for solving the interval analysis based problems \cite{Ru99a}.  




\section{Applications}
\label{sec:application}
Context serves to demonstrate the applications of the aforementioned uncertainty modeling techniques. 
The applications are widely categorized into several fields, as given in Table \ref{tab:application}. The summaries of uncertainty modeling attributes are provided in Table \ref{tab:attrib}. 

\begin{itemize}
	\item {DG impact assessment}
	\item {Plug-in hybrid electric vehicle (PHEV): (e.g. exploitation of plug in hybrid electric vehicles)}
	\item {Assessment of available transfer capability (ATC)}
	\item {Renewable energy (operation and planning)(e.g. hydro power generation management )}
	\item {Load flow/optimal power flow calculations (e.g. probabilistic load flow, fuzzy load flow.)}
	\item {Reliability evaluation (e.g. reliability-oriented distribution network reconfiguration)}
	\item {Distribution network operation and planning (e.g. phase balancing, cost-benefit analysis of distribution automation)}
	\item {Transmission/Generation planning, operation and control: (e.q. self-scheduling of gencos, fault location scheme, dynamic economic dispatch, maintenance scheduling, determination of pilot points for zonal voltage control, small-signal stability )}
	\item{State estimation} 
	\item{Electricity market (e.g. real time demand side management, bidding strategy,energy hub management and electricity procurement strategy.)} 
\end{itemize}

\section{Promising lines of future researches}
\label{sec:trends}
The future trends in uncertainty modeling \textcolor{black}{to investigate and further explore} are summarized as follows:
\subsection{Exploring new uncertain parameters}

With the increasingly revolutionary changes in power system's regulatory framework and developing technologies the uncertainty in input data of decision making procedures is increased. These uncertain environment include financial, societal/governmental (the ongoing government policy and the future potential incentive for the renewable energy), environmental (carbon emission and global warming issue) and technical (communication and information architecture in smart grids, demand response, PHEV, energy hubs, smart building) uncertainties, risk preferences in the investment models, fuel prices and market regulations, renewable energy sources and competition among suppliers. 

\subsection{Enhancing the existing techniques}
\begin{itemize}
	\item Reduce the computational burden specially when applied to large scale power systems and real-time applications
	\item Choosing the appropriate uncertainty handling technique
	\item Hybridizing the existing techniques to better describe the uncertain environment
	\item Using the heuristic methods to soften the computation procedures
\end{itemize}

\subsection{Exploring the new uncertainty handling methods}
The taxonomy of the uncertainty modeling methods in past, present and future is as depicted in Fig.\ref{fig:trend}. In 2011, Zadeh introduced a new class of uncertain numbers called ``\textit{Z-numbers}'' \cite{Zadeh20112923}. The Z-numbers are expressed as a pair in form of $Z=(A,B)$, in which, $A,B$ are restrictions describing the behavior of $Z$.
$A$ is usually a fuzzy set while $B$ describes the certainty degree. The certainty degree may be expressed as a PDF or a fuzzy set. In this context, $Z= \left\{x|x \in A \mbox{ with certainty degree equal to } B\right\}$. In classic fuzzy numbers decision maker just has $A$ and it is quit sure that $Z$ belongs to $A$. However in Z-numbers, $Z$ is described using the set $A$ with a certainty (reliability) degree of information called $B$. Examples for Z-numbers are provided in Table \ref{tab:example}. 

 The normal PDF, as a function of $\mu,\sigma$, is a reasonable choice for modeling the randomness of the load variable.          
In order to disambiguate this concept, a simple two-bus network is used as shown in Fig.\ref{fig:network}.
The value of load can be described in various way as described in Table \ref{tab:zload}. 
For example, we are \textit{almost certain} (set $B_2$) that \textit{the demand value} in a given bus ($Z$ number) is \textit{low} (set $A_1$) as depicted in Fig.\ref{fig:shekl2}. The probability that \textit{the load value} is \textit{low} can be calculated as (\ref{eq:znumber}). 
\begin{figure*}
\begin{alignat}{2}
\label{eq:znumber}
&Prob=\int^d_a A_1\frac{1}{\sigma \sqrt{2\pi}}e^{\frac{-(x-\mu)^2}{2\sigma^2}}=\frac{1}{\sigma \sqrt{2\pi}}[\int^b_a \frac{x-a}{b-a}e^{\frac{-(x-\mu)^2}{2\sigma^2}} + \int^c_b e^{\frac{-(x-\mu)^2}{2\sigma^2}}+ \int^d_c \frac{x-d}{c-d}e^{\frac{-(x-\mu)^2}{2\sigma^2}}] \\
&G(Prob)=\mu_{B_2}(Prob)  
\end{alignat}
\end{figure*}
In (\ref{eq:znumber}), $G(Prob)$ indicates the degree to which $Prob$ belongs to $A_1$. 
Now, the information of Z-number expressed as $L=(A_1,B_2)$ for load parameter is represented as a possibility distribution ($G(Prob)$) over the space of probability distributions (various values of $\mu,\sigma$). 

	
%
\section{Conclusion}
\label{sec:conclu}
This paper proposed a standard classification of uncertainty handling methods along with the promising lines of future researches. The possibility of using Z-numbers for uncertainty modeling of load values was introduced for the first time.
The assessed methodologies include probabilistic, possibilistic, hybrid methods, robust optimization, interval based analysis as well as Z-numbers. These models are compared and their strength and shortcomings are investigated.
Based on the proposed comprehensive classification, it is deduced that each method is suitable for a specific type of uncertainty. The severity of uncertainty dictates choosing the appropriate uncertainty modeling technique. Additionally, according to the carried out taxonomy of the methodologies, it was revealed that some research areas are still remained untouched.

\bibliographystyle{elsarticle-num}
\bibliography{ref}

\begin{thebibliography}{10}
\expandafter\ifx\csname url\endcsname\relax
  \def\url#1{\texttt{#1}}\fi
\expandafter\ifx\csname urlprefix\endcsname\relax\def\urlprefix{URL }\fi
\expandafter\ifx\csname href\endcsname\relax
  \def\href#1#2{#2} \def\path#1{#1}\fi

\bibitem{attoh2005applied}
N.~Attoh-Okine, B.~Ayyub, Applied research in uncertainty modeling and
  analysis, Vol.~20, Springer Verlag, 2005.

\bibitem{conejou}
A.~J. Conejo, M.~Carrion, J.~M. Morales, Decision Making Under Uncertainty in
  Electricity Markets, Springer, New York, 2010.

\bibitem{dantzig2011linear}
G.~Dantzig, Linear programming under uncertainty, Stochastic Programming (2011)
  1--11.

\bibitem{zadeh1965fuzzy}
L.~Zadeh, Fuzzy sets, Information and control 8~(3) (1965) 338--353.

\bibitem{IGDT}
Y.~B. Haim, Info-Gap Decision Theory (Second Edition), Academic Press,
  California, 2006.

\bibitem{Soyster}
A.~L. Soyster, Convex programming with set-inclusive constraints and
  applications to inexact linear programming, Journal of Operations Research
  21~(2) (1973) 1154--1157.

\bibitem{moore1966interval}
R.~Moore, Interval analysis, Prentice-Hall, Englewood Cliff, New Jersey, 1966.

\bibitem{mcs}
M.~H. Kalos, P.~A. Whitlock, Monte Carlo Methods, WILEY-VCH Verlag GmbH and Co.
  KGaA, 2004.

\bibitem{4808223}
H.~Yu, C.~Chung, K.~Wong, H.~Lee, J.~Zhang, Probabilistic load flow evaluation
  with hybrid latin hypercube sampling and cholesky decomposition, IEEE
  Transactions on Power Systems, 24~(2) (2009) 661 --667.

\bibitem{Hong1998261}
H.~P. Hong, An efficient point estimate method for probabilistic analysis,
  Reliability Engineering and System Safety 59~(3) (1998) 261 -- 267.

\bibitem{4801580}
J.~Morales, S.~Pineda, A.~Conejo, M.~Carrion, Scenario reduction for futures
  market trading in electricity markets, IEEE Trans. on Power Sys., 24~(2)
  (2009) 878 --888.

\bibitem{5473192}
S.~Pineda, A.~Conejo, Scenario reduction for risk-averse electricity trading,
  Generation, Transmission Distribution, IET 4~(6) (2010) 694 --705.

\bibitem{6128764}
T.~Amraee, A.~Soroudi, A.~Ranjbar, Probabilistic determination of pilot points
  for zonal voltage control, Generation, Transmission Distribution, IET 6~(1)
  (2012) 1 --10.

\bibitem{466473}
J.~Momoh, X.~Ma, K.~Tomsovic, Overview and literature survey of fuzzy set
  theory in power systems, IEEE Transactions on Power Systems, 10~(3) (1995)
  1676 --1690.

\bibitem{fuzzycontrol}
H.~Zhang, D.~Liu (Eds.), Fuzzy Modeling and Fuzzy Control, Birkhäuser, 2006.

\bibitem{deffuz}
T.~Ross (Ed.), Fuzzy logic with engineering applications, Wiley, 2004.

\bibitem{Soroudi2011794}
A.~Soroudi, M.~Ehsan, A possibilistic–probabilistic tool for evaluating the
  impact of stochastic renewable and controllable power generation on energy
  losses in distribution networks—a case study, Renewable and Sustainable
  Energy Reviews 15~(1) (2011) 794 -- 800.

\bibitem{6142135}
A.~Soroudi, Possibilistic-scenario model for dg impact assessment on
  distribution networks in an uncertain environment, IEEE Transactions on Power
  Systems, 27~(3) (2012) 1283 -- 1293.

\bibitem{6338333}
A.~Soroudi, M.~Ehsan, Igdt based robust decision making tool for dnos in load
  procurement under severe uncertainty, IEEE Transactions on Power Systems,
  PP~(99) (2013) 1--10.

\bibitem{6298065}
B.~Mohammadi-Ivatloo, H.~Zareipour, N.~Amjady, M.~Ehsan, Application of
  information-gap decision theory to risk-constrained self-scheduling of
  gencos, IEEE Transactions on Power Systems, PP~(99) (2012) 1--10.

\bibitem{bental}
A.~Ben-Tal, L.~E. Ghaoui, A.~Nemirovski, Robust Optimization, 1st Edition,
  Princeton Series in Applied Mathematics, 2009.

\bibitem{priceofro}
D.~Bertsimas, M.~Sim, The price of robustness, Journal of Operations Research
  52~(1) (2004) 35--53.

\bibitem{rosoft}
M.~S. Joel~Goh, {Robust Optimization Made Easy with ROME}, Vol.~59, Informs,
  Dordrecht, 2011, p. 973–985.

\bibitem{Ru99a}
S.~Rump, {INTLAB - INTerval LABoratory}, in: T.~Csendes (Ed.),
  {Developments~in~Reliable Computing}, Kluwer Academic Publishers, Dordrecht,
  1999, pp. 77--104, \url{http://www.ti3.tu-harburg.de/rump/}.

\bibitem{Zadeh20112923}
L.~A. Zadeh, A note on z-numbers, Information Sciences 181~(14) (2011) 2923 --
  2932.

\bibitem{1626355}
W.~El-Khattam, Y.~Hegazy, M.~Salama, Investigating distributed generation
  systems performance using monte carlo simulation, IEEE Transactions on Power
  Systems, 21~(2) (2006) 524 -- 532.

\bibitem{6074999}
A.~Soroudi, R.~Caire, N.~Hadjsaid, M.~Ehsan, Probabilistic dynamic
  multi-objective model for renewable and non-renewable distributed generation
  planning, Generation, Transmission Distribution, IET 5~(11) (2011) 1173
  --1182.

\bibitem{5340599}
C.-L. Su, Stochastic evaluation of voltages in distribution networks with
  distributed generation using detailed distribution operation models, IEEE
  Transactions on Power Systems, 25~(2) (2010) 786 --795.

\bibitem{6026244}
Z.~Liu, F.~Wen, G.~Ledwich, Optimal siting and sizing of distributed generators
  in distribution systems considering uncertainties, IEEE Transactions on Power
  Delivery, 26~(4) (2011) 2541 --2551.

\bibitem{5743048}
V.~Martins, C.~Borges, Active distribution network integrated planning
  incorporating distributed generation and load response uncertainties, IEEE
  Transactions on Power Systems, 26~(4) (2011) 2164 --2172.

\bibitem{5733385}
A.~Soroudi, M.~Ehsan, R.~Caire, N.~Hadjsaid, Possibilistic evaluation of
  distributed generations impacts on distribution networks, IEEE Transactions
  on Power Systems, 26~(4) (2011) 2293 --2301.

\bibitem{6193193}
A.~Lojowska, D.~Kurowicka, G.~Papaefthymiou, L.~van~der Sluis, Stochastic
  modeling of power demand due to evs using copula, IEEE Transactions on Power
  Systems, PP~(99) (2012) 1.

\bibitem{6069832}
M.~Pantos, Exploitation of electric-drive vehicles in electricity markets, IEEE
  Transactions on Power Systems, 27~(2) (2012) 682 --694.

\bibitem{5720537}
A.~Hajimiragha, C.~Canizares, M.~Fowler, S.~Moazeni, A.~Elkamel, A robust
  optimization approach for planning the transition to plug-in hybrid electric
  vehicles, IEEE Transactions on Power Systems, 26~(4) (2011) 2264 --2274.

\bibitem{4077137}
A.~B. Rodrigues, M.~G. Da~Silva, Probabilistic assessment of available transfer
  capability based on monte carlo method with sequential simulation, IEEE
  Transactions on Power Systems, 22~(1) (2007) 484 --492.

\bibitem{1425547}
C.-L. Su, C.-N. Lu, Two-point estimate method for quantifying transfer
  capability uncertainty, IEEE Transactions on Power Systems, 20~(2) (2005) 573
  -- 579.

\bibitem{1295027}
A.~Khairuddin, S.~Ahmed, M.~Mustafa, A.~Zin, H.~Ahmad, A novel method for atc
  computations in a large-scale power system, IEEE Transactions on Power
  Systems, 19~(2) (2004) 1150 -- 1158.

\bibitem{6084851}
S.~Bu, W.~Du, H.~Wang, Z.~Chen, L.~Xiao, H.~Li, Probabilistic analysis of
  small-signal stability of large-scale power systems as affected by
  penetration of wind generation, IEEE Transactions on Power Systems, 27~(2)
  (2012) 762 --770.

\bibitem{5967923}
Q.~Wang, Y.~Guan, J.~Wang, A chance-constrained two-stage stochastic program
  for unit commitment with uncertain wind power output, IEEE Transactions on
  Power Systems, 27~(1) (2012) 206 --215.

\bibitem{6003751}
A.~Soroudi, M.~Aien, M.~Ehsan, A probabilistic modeling of photo voltaic
  modules and wind power generation impact on distribution networks, Systems
  Journal, IEEE 6~(2) (2012) 254 --259.

\bibitem{496139}
L.~Escudero, J.~de~la Fuente, C.~Garcia, F.~Prieto, Hydropower generation
  management under uncertainty via scenario analysis and parallel computation,
  IEEE Transactions on Power Systems, 11~(2) (1996) 683 --689.

\bibitem{5929570}
Y.~Makarov, P.~Etingov, J.~Ma, Z.~Huang, K.~Subbarao, Incorporating uncertainty
  of wind power generation forecast into power system operation, dispatch, and
  unit commitment procedures, IEEE Transactions on Sustainable Energy 2~(4)
  (2011) 433 --442.

\bibitem{6007045}
A.~Saber, G.~Venayagamoorthy, Resource scheduling under uncertainty in a smart
  grid with renewables and plug-in vehicles, Systems Journal, IEEE 6~(1) (2012)
  103 --109.

\bibitem{5734883}
C.~Baslis, A.~Bakirtzis, Mid-term stochastic scheduling of a price-maker hydro
  producer with pumped storage, IEEE Transactions on Power Systems, 26~(4)
  (2011) 1856 --1865.

\bibitem{5986768}
S.~Rebennack, B.~Flach, M.~Pereira, P.~Pardalos, Stochastic hydro-thermal
  scheduling under emissions constraints, IEEE Transactions on Power Systems,
  27~(1) (2012) 58 --68.

\bibitem{4636752}
B.~Venkatesh, P.~Yu, H.~Gooi, D.~Choling, Fuzzy milp unit commitment
  incorporating wind generators, IEEE Transactions on Power Systems, 23~(4)
  (2008) 1738 --1746.

\bibitem{6069602}
R.~Jiang, J.~Wang, Y.~Guan, Robust unit commitment with wind power and pumped
  storage hydro, IEEE Transactions on Power Systems, PP~(99) (2011) 1.

\bibitem{5462528}
H.~Zhang, P.~Li, Probabilistic analysis for optimal power flow under
  uncertainty, Generation, Transmission Distribution, IET 4~(5) (2010) 553
  --561.

\bibitem{4567448}
X.~Li, Y.~Li, S.~Zhang, Analysis of probabilistic optimal power flow taking
  account of the variation of load power, IEEE Transactions on Power Systems,
  23~(3) (2008) 992 --999.

\bibitem{Gouveia20091012}
E.~M. Gouveia, M.~A. Matos, Symmetric ac fuzzy power flow model, European
  Journal of Operational Research 197~(3) (2009) 1012 -- 1018.

\bibitem{4435948}
M.~Matos, E.~Gouveia, The fuzzy power flow revisited, IEEE Transactions on
  Power Systems, 23~(1) (2008) 213 --218.

\bibitem{5739617}
A.~Romero, H.~Zini, G.~Ratta?~and, R.~Dib, Harmonic load-flow approach based on
  the possibility theory, Generation, Transmission Distribution, IET 5~(4)
  (2011) 393 --404.

\bibitem{207353}
Z.~Wang, F.~Alvarado, Interval arithmetic in power flow analysis, IEEE
  Transactions on Power Systems, 7~(3) (1992) 1341 --1349.

\bibitem{5325712}
R.~Karki, P.~Hu, R.~Billinton, Reliability evaluation considering wind and
  hydro power coordination, IEEE Transactions on Power Systems, 25~(2) (2010)
  685 --693.

\bibitem{5416324}
L.~Wu, M.~Shahidehpour, Y.~Fu, Security-constrained generation and transmission
  outage scheduling with uncertainties, IEEE Transactions on Power Systems,
  25~(3) (2010) 1674 --1685.

\bibitem{5739566}
A.~Papavasiliou, S.~Oren, R.~O'Neill, Reserve requirements for wind power
  integration: A scenario-based stochastic programming framework, IEEE
  Transactions on Power Systems, 26~(4) (2011) 2197 --2206.

\bibitem{4162625}
L.~Wu, M.~Shahidehpour, T.~Li, Stochastic security-constrained unit commitment,
  IEEE Transactions on Power Systems, 22~(2) (2007) 800 --811.

\bibitem{5405057}
H.~Ge, S.~Asgarpoor, Reliability evaluation of equipment and substations with
  fuzzy markov processes, IEEE Transactions on Power Systems, 25~(3) (2010)
  1319 --1328.

\bibitem{Zhang2012138}
P.~Zhang, W.~Li, S.~Wang, Reliability-oriented distribution network
  reconfiguration considering uncertainties of data by interval analysis,
  International Journal of Electrical Power \& Energy Systems 34~(1) (2012) 138
  -- 144.

\bibitem{5686893}
C.~yang Li, X.~Chen, X.~shan Yi, J.~yong Tao, Interval-valued reliability
  analysis of multi-state systems,, Reliability, IEEE Transactions on 60~(1)
  (2011) 323 --330.

\bibitem{847282}
P.~Carvalho, L.~Ferreira, F.~Lobo, L.~Barruncho, Distribution network expansion
  planning under uncertainty: a hedging algorithm in an evolutionary approach,
  IEEE Transactions on Power Delivery, 15~(1) (2000) 412 --416.

\bibitem{1350817}
I.~Ramirez-Rosado, J.~Dominguez-Navarro, Possibilistic model based on fuzzy
  sets for the multiobjective optimal planning of electric power distribution
  networks, IEEE Transactions on Power Systems, 19~(4) (2004) 1801 -- 1810.

\bibitem{Shaalan1993145}
H.~E. Shaalan, R.~P. Broadwater, Using interval mathematics in cost-benefit
  analysis of distribution automation, Electric Power Systems Research 27~(2)
  (1993) 145 -- 152.

\bibitem{5382489}
A.~Kazerooni, J.~Mutale, Transmission network planning under security and
  environmental constraints, IEEE Transactions on Power Systems, 25~(2) (2010)
  1169 --1178.

\bibitem{5937035}
S.~Kazempour, A.~Conejo, Strategic generation investment under uncertainty via
  benders decomposition, IEEE Transactions on Power Systems, 27~(1) (2012) 424
  --432.

\bibitem{4282038}
T.~Li, M.~Shahidehpour, Risk-constrained generation asset arbitrage in power
  systems,, IEEE Transactions on Power Systems, 22~(3) (2007) 1330 --1339.

\bibitem{1490592}
A.~Saric, A.~Stankovic, Model uncertainty in security assessment of power
  systems,, IEEE Transactions on Power Systems, 20~(3) (2005) 1398 -- 1407.

\bibitem{1388546}
H.~Yamin, Fuzzy self-scheduling for gencos, IEEE Transactions on Power Systems,
  20~(1) (2005) 503 -- 505.

\bibitem{1318664}
P.~Attaviriyanupap, H.~Kita, E.~Tanaka, J.~Hasegawa, A fuzzy-optimization
  approach to dynamic economic dispatch considering uncertainties, IEEE
  Transactions on Power Systems, 19~(3) (2004) 1299 -- 1307.

\bibitem{Mohanta200473}
D.~K. Mohanta, P.~K. Sadhu, R.~Chakrabarti, Fuzzy reliability evaluation of
  captive power plant maintenance scheduling incorporating uncertain forced
  outage rate and load representation, Electric Power Systems Research 72~(1)
  (2004) 73 -- 84.

\bibitem{6070998}
Y.-Y. Hong, P.-H. Chen, Genetic-based underfrequency load shedding in a
  stand-alone power system considering fuzzy loads, Power Delivery, IEEE
  Transactions on 27~(1) (2012) 87 --95.

\bibitem{4745920}
Z.~Xu, M.~Liu, G.~Yang, N.~Li, Application of interval analysis and evidence
  theory to fault location, Electric Power Applications, IET 3~(1) (2009) 77
  --84.

\bibitem{Zhu201271}
Y.~Zhu, Y.~Li, G.~Huang, Planning municipal-scale energy systems under
  functional interval uncertainties, Renewable Energy 39~(1) (2012) 71 -- 84.

\bibitem{5710446}
Y.~Wang, Q.~Xia, C.~Kang, Unit commitment with volatile node injections by
  using interval optimization, IEEE Transactions on Power Systems, 26~(3)
  (2011) 1705 --1713.

\bibitem{5659501}
A.~Street, F.~Oliveira, J.~Arroyo, Contingency-constrained unit commitment with
  n - k security criterion: A robust optimization approach, IEEE Transactions
  on Power Systems, 26~(3) (2011) 1581 --1590.

\bibitem{4810103}
R.~Singh, B.~Pal, R.~Vinter, Measurement placement in distribution system state
  estimation, IEEE Transactions on Power Systems, 24~(2) (2009) 668 --675.

\bibitem{5438873}
E.~Caro, J.-M. Morales, A.~Conejo, R.~Minguez, Calculation of measurement
  correlations using point estimate, IEEE Transactions on Power Delivery,
  25~(4) (2010) 2095 --2103.

\bibitem{982205}
K.-R. Shih, S.-J. Huang, Application of a robust algorithm for dynamic state
  estimation of a power system, IEEE Transactions on Power Systems, 17~(1)
  (2002) 141 --147.

\bibitem{1193880}
A.~Saric, R.~Ciric, Integrated fuzzy state estimation and load flow analysis in
  distribution networks, IEEE Transactions on Power Delivery, 18~(2) (2003) 571
  -- 578.

\bibitem{6153413}
Y.~Wang, W.~Li, P.~Zhang, B.~Wang, J.~Lu, Reliability analysis of phasor
  measurement unit considering data uncertainty, IEEE Transactions on Power
  Systems, PP~(99) (2012) 1 --8.

\bibitem{5999751}
C.~Rakpenthai, S.~Uatrongjit, S.~Premrudeeprechacharn, State estimation of
  power system considering network parameter uncertainty based on parametric
  interval linear systems,, IEEE Transactions on Power Systems, 27~(1) (2012)
  305 --313.

\bibitem{5986767}
M.~Perninge, F.~Lindskog, L.~Soder, Importance sampling of injected powers for
  electric power system security analysis, IEEE Transactions on Power Systems,
  27~(1) (2012) 3 --11.

\bibitem{5714772}
P.~Duenas, J.~Reneses, J.~Barquin, Dealing with multi-factor uncertainty in
  electricity markets by combining monte carlo simulation with spatial
  interpolation techniques, Generation, Transmission Distribution, IET 5~(3)
  (2011) 323 --331.

\bibitem{1626354}
A.~Saric, A.~Stankovic, An application of interval analysis and optimization to
  electric energy markets, IEEE Transactions on Power Systems, 21~(2) (2006)
  515 -- 523.

\bibitem{5607339}
A.~Conejo, J.~Morales, L.~Baringo, Real-time demand response model, IEEE
  Transactions on Smart Grid 1~(3) (2010) 236 --242.

\bibitem{1378725}
M.-P. Cheong, D.~Berleant, G.~Sheble, Information gap decision theory as a tool
  for strategic bidding in competitive electricity markets, in: Probabilistic
  Methods Applied to Power Systems, 2004 International Conference on, 2004, pp.
  421 --426.

\bibitem{Parisio201298}
A.~Parisio, C.~D. Vecchio, A.~Vaccaro, A robust optimization approach to energy
  hub management, International Journal of Electrical Power \& Energy Systems
  42~(1) (2012) 98 -- 104.

\bibitem{5723039}
K.~Zare, M.~Moghaddam, M.~Sheikh-El-Eslami, Risk-based electricity procurement
  for large consumers, IEEE Transactions on Power Systems, 26~(4) (2011) 1826
  --1835.

\end{thebibliography}

\clearpage
\newpage
List of Figure Captions:
\begin{itemize}
	\item Figure 1. General classification of uncertain parameters in energy system studies
	\item Figure 2. Uncertainty modeling tools
	\item Figure 3. Uncertainty modeling trends: past, present and future
	\item Figure 4. Simple two-bus illustrative network
  \item Figure 5. Concept of Z-number 
\end{itemize}
\newpage
\begin{table*} 
\centering
\caption{Summaries of uncertainty modeling applications}
\begin{threeparttable}
	\scalebox{0.5}{
		\begin{tabular}{@{}m{7cm}|l|l|l|l|l|l|l|l|@{}}
\toprule[1.5pt]
\textbf{Applications}	&	\multicolumn{3}{|c|}{\textbf{Probabilistic}}	&	\textbf{Possibilistic}	&	\textbf{Hybrid} 	&	\textbf{Interval}	&\textbf{RO}	&	\textbf{IDGT}	\\ \toprule[1.5pt]
	&	MC	&PEM	&	Scenario	& 	\multicolumn{5}{c}{} \\ \toprule[1.5pt]
\multicolumn{1}{|l|}{DG units }	&	\cite{1626355,6074999}	&	\cite{5340599}	&	\cite{6026244,5743048}	&	\multicolumn{1}{c|}{\cite{5733385}}	&	\cite{Soroudi2011794,6142135}	&	\cellcolor{orange!30} \tnote{\textdagger}	&	\cellcolor{orange!30}	&	\cellcolor{orange!30}	\\ \hline
\multicolumn{1}{|l|}{PHEV}	&	\cite{6026244,6193193}	&	\cite{6069832}	&	\cite{6069832}	&	\cellcolor{orange!30}	&	\cellcolor{orange!30}	&	\cellcolor{orange!30}	&	\cite{5720537}	&	\cellcolor{orange!30}	\\\hline
\multicolumn{1}{|l|}{Available transfer capability (ATC)} 	&	\cite{4077137}	&	\cite{1425547}	&	\cellcolor{orange!30}	&	\cite{1295027}	&	\cellcolor{orange!30}	&	\cellcolor{orange!30}	&	\cellcolor{orange!30}	&	\cellcolor{orange!30}	\\\hline
\multicolumn{1}{|l|}{Renewable energy (operation and planning)} 	&	\cite{6084851,5967923}	&	\cite{6003751}	&	 \cite{496139,5929570,6007045,5734883,5986768} 	&	\cite{4636752}	&	\cellcolor{orange!30}	&	\cellcolor{orange!30}	&	 \cite{6069602}	&	\cellcolor{orange!30}	\\\hline
\multicolumn{1}{|l|}{Load flow/Optimal power flow} 	&	\cite{5462528}	&	\cite{4567448}	&	\cellcolor{orange!30}	&	\cite{Gouveia20091012,4435948,5739617}	&	\cellcolor{orange!30}	&	\cite{207353}	&	\cellcolor{orange!30}	&	\cellcolor{orange!30}	\\\hline
\multicolumn{1}{|l|}{Reliability evaluation}	&	\cite{5325712,5416324}	&	\cite{5739566}	&	 \cite{4162625,5416324}	&	\cite{5405057}	&	\cellcolor{orange!30}	&	\cite{Zhang2012138,5686893}	&		\cellcolor{orange!30} &	\cellcolor{orange!30}	\\ \hline
\multicolumn{1}{|l|}{Distribution operation and planning}	&	\cite{6074999}	&	\cellcolor{orange!30}	&	\cite{847282}	&	\cite{1350817}	&		&	\cite{Shaalan1993145}	&	\cellcolor{orange!30}	&	\cellcolor{orange!30}	\\\hline
\multicolumn{1}{|l|}{Transmission/generation planning and operation/control}	&	\cite{5382489}	&	\cellcolor{orange!30}	&	\cite{5937035,4282038,6128764,1490592}	&	 \cite{1388546,1318664,Mohanta200473,6070998}	&	\cite{1318664}	&	\cite{4745920,Zhu201271,5710446}	&	\cite{5659501}	&	\cellcolor{orange!30}	\\\hline
\multicolumn{1}{|l|}{State estimation}	&	\cite{4810103}	&	 \cite{5438873}	&	\cite{982205}	&	\cite{1193880,6153413}	&	\cite{1193880}	&	\cite{5999751}	&	\cellcolor{orange!30}	&	\cellcolor{orange!30}	\\\hline
\multicolumn{1}{|l|}{Electricity market}	&	\cite{5986767,5714772}	&	\cellcolor{orange!30}	&	\cellcolor{orange!30}	&	\multicolumn{1}{c|}{\cite{1388546}}	&	\multicolumn{1}{c|}{\cellcolor{orange!30}}	&	\cite{1626354}	&	\cite{5607339,1378725,Parisio201298}	&	\cite{5723039,1378725}	\\
\bottomrule[1.5pt]
\end{tabular}}
\begin{tablenotes}
\item[\textdagger] Unexplored research directions
\end{tablenotes}
\end{threeparttable}
\label{tab:application}
\end{table*} 
------

\begin{table*}[ht]
	\centering
	\caption{Summaries of uncertainty modeling attributes}
	\rowcolors{1}{white}{orange!30}
	\scalebox{0.8}{
		\begin{tabular}{@{}m{2.3cm}m{2.5cm}m{4.1cm}m{3.8cm}m{4cm}m{4cm}|@{}}
\toprule[1.5pt]
\textbf{Method} 	&	\textbf{Input representation}	&	\textbf{Output attributes}	&	\textbf{Advantages}	&	\textbf{Disadvantages}	\\  
\toprule[1.5pt]
\textsl{Probabilistic}	&	PDF	&	Statistics like expectation, variance, etc.	&	Easy to implement 	&	Computationally expensive, needs a large amount of historic data, approximate result	\\ \hline
\textsl{Possibilistic} 	&	MF	&	MF	&	Converting linguistic knowledge to numerical values	&	Complex implementation	\\ \hline
\textsl{Hybrid} 	&	 MF \& PDF	&	Membership function with probabilistic parameters	&	Dealing with both uncertainty types simultaneously	&	Computationally expensive	\\ \hline
\textit{IGDT}	&	Forecasted values 	&	Decision variables satisfying the requirements 	&	Useful for severe uncertainties	&	Too conservative 	\\ \hline
\textsl{Robust Optimization	}&	Intervals 	&	Controlled conservativeness	&	Useful when just an interval is available 	&	Difficult to use in non-linear models 	\\ \hline
\textsl{Interval Analysis}	&	Intervals 	&	Bounds of the outputs	&	Useful when just an interval is available 	&	The correlations among intervals are neglected this would make it too conservative	\\
\bottomrule[1.5pt]
		\end{tabular}}
	\label{tab:attrib}
\end{table*} 

\begin{table}[ht]
	\centering
	\caption{Examples for Z-numbers}
	\scalebox{0.9}{
		\begin{tabular}{lll}
\toprule[1.5pt]
\textbf{Parameter} 	&	\textbf{A}	&	\textbf{B}		\\  
\toprule[1.5pt]
Demand value        & High        & Very sure     \\
Wind speed          & Weibul PDF  & Normally      \\
Voltage magnitude   & Uniform distribution in $[0.95 1.05]$ & In most cases \\
\bottomrule[1.5pt]
		\end{tabular}}
	\label{tab:example}
\end{table} 

\begin{table}[ht]
	\centering
	\caption{Describing the load values as Z-numbers}
	\scalebox{0.8}{
		\begin{tabular}{clc}
\toprule[1.5pt]
	\multicolumn{1}{c}{\textbf{A}}	&	\multicolumn{1}{c}{\textbf{B}} & \textbf{Load}		\\  
\toprule[1.5pt]
\multirow{3}{*}{Low} 
  & Not sure &	$L=(A_1,B_1)$ \\ 
  & Almost certain &	$L=(A_1,B_2)$ \\ 
  & Quit sure &	$L=(A_1,B_3)$ \\ \hline
\multirow{3}{*}{Medium} 
  & Not sure &	$L=(A_2,B_1)$ \\ 
  & Almost certain &	$L=(A_2,B_2)$ \\ 
  & Quit sure &	$L=(A_2,B_3)$ \\ \hline
  \multirow{3}{*}{High} 
  & Not sure &	$L=(A_3,B_1)$ \\ 
  & Almost certain &	$L=(A_3,B_2)$ \\ 
  & Quit sure &	$L=(A_3,B_3)$ \\
\bottomrule[1.5pt]
		\end{tabular}}
	\label{tab:zload}
\end{table} 

\newpage
\clearpage
\begin{figure}[ht]
	\centering
		\includegraphics[clip = true,width=1\columnwidth,bb=10 350 800 780]{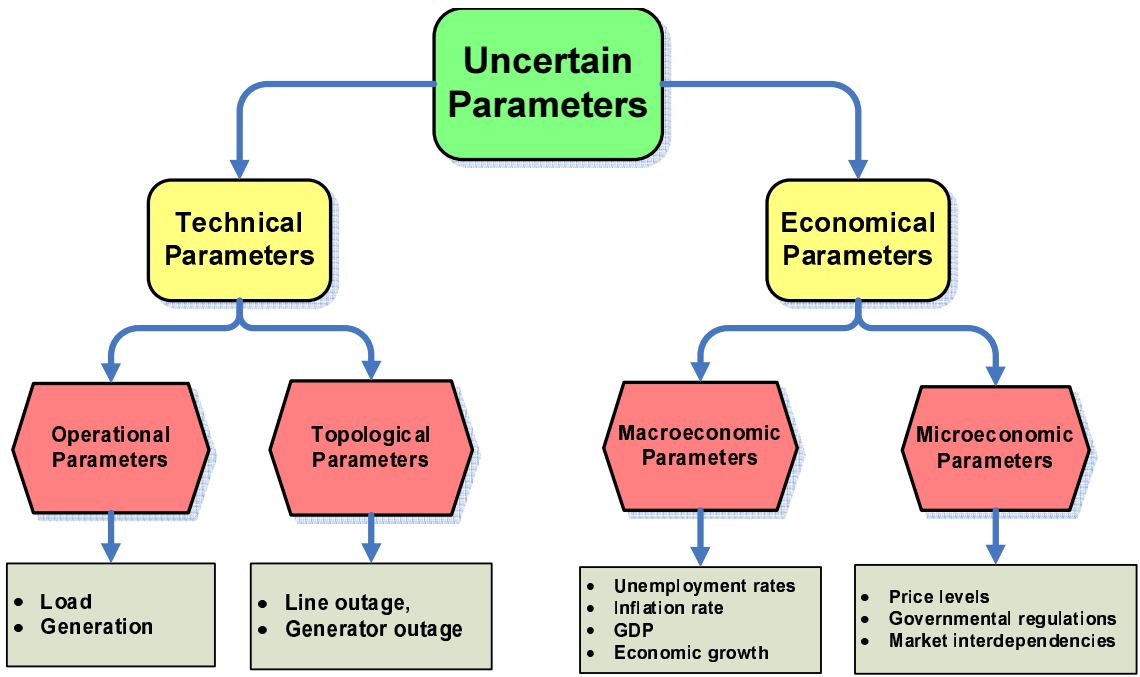}
	\caption{General classification of uncertain parameters in energy system studies}
	\label{fig:flow}
\end{figure}

\begin{figure}[ht]
	\centering
		\includegraphics[clip = true,width=1\columnwidth,bb=10 420 830 800]{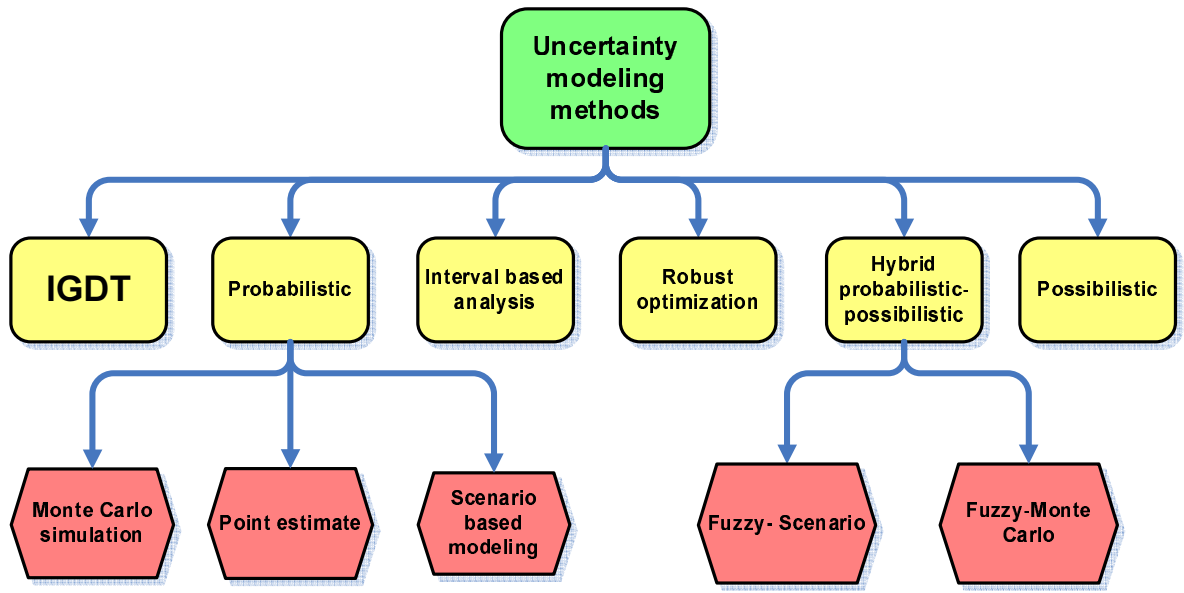}
	\caption{Uncertainty modeling tools}
	\label{fig:methods}
\end{figure}
\begin{figure}[ht]
	\centering
		\includegraphics[clip = true,width=0.6\columnwidth,bb=100 250 500 650]{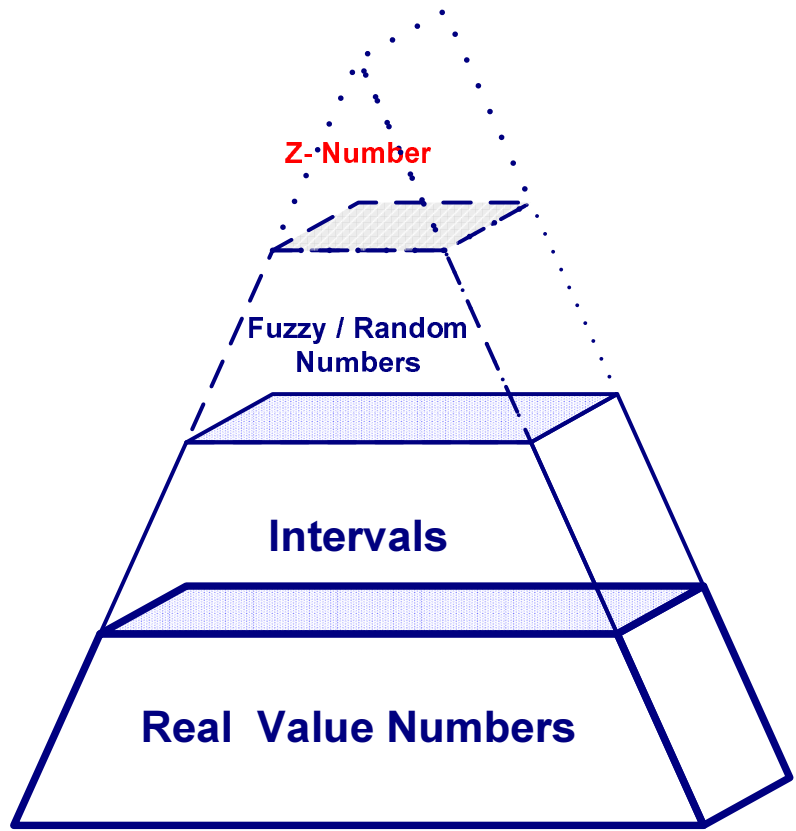}
	\caption{Uncertainty modeling trends: past, present and future}
	\label{fig:trend}
\end{figure}

\begin{figure}[ht]
	\centering
		\includegraphics[clip = true,width=0.6\columnwidth,bb=100 350 500 530]{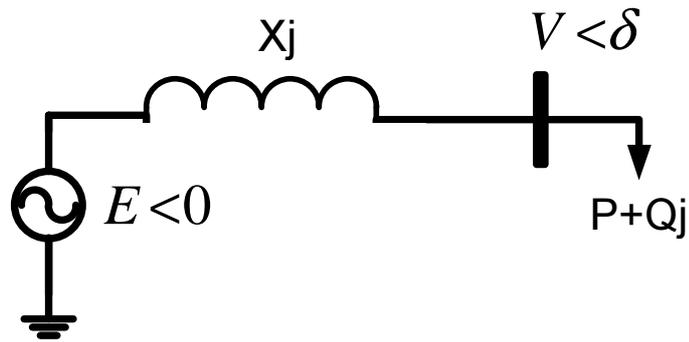}
	\caption{Simple two-bus illustrative network}
	\label{fig:network}
\end{figure}

\begin{figure}[ht]
	\centering
		\includegraphics[clip = true,width=1\columnwidth,bb=120 160 600 600]{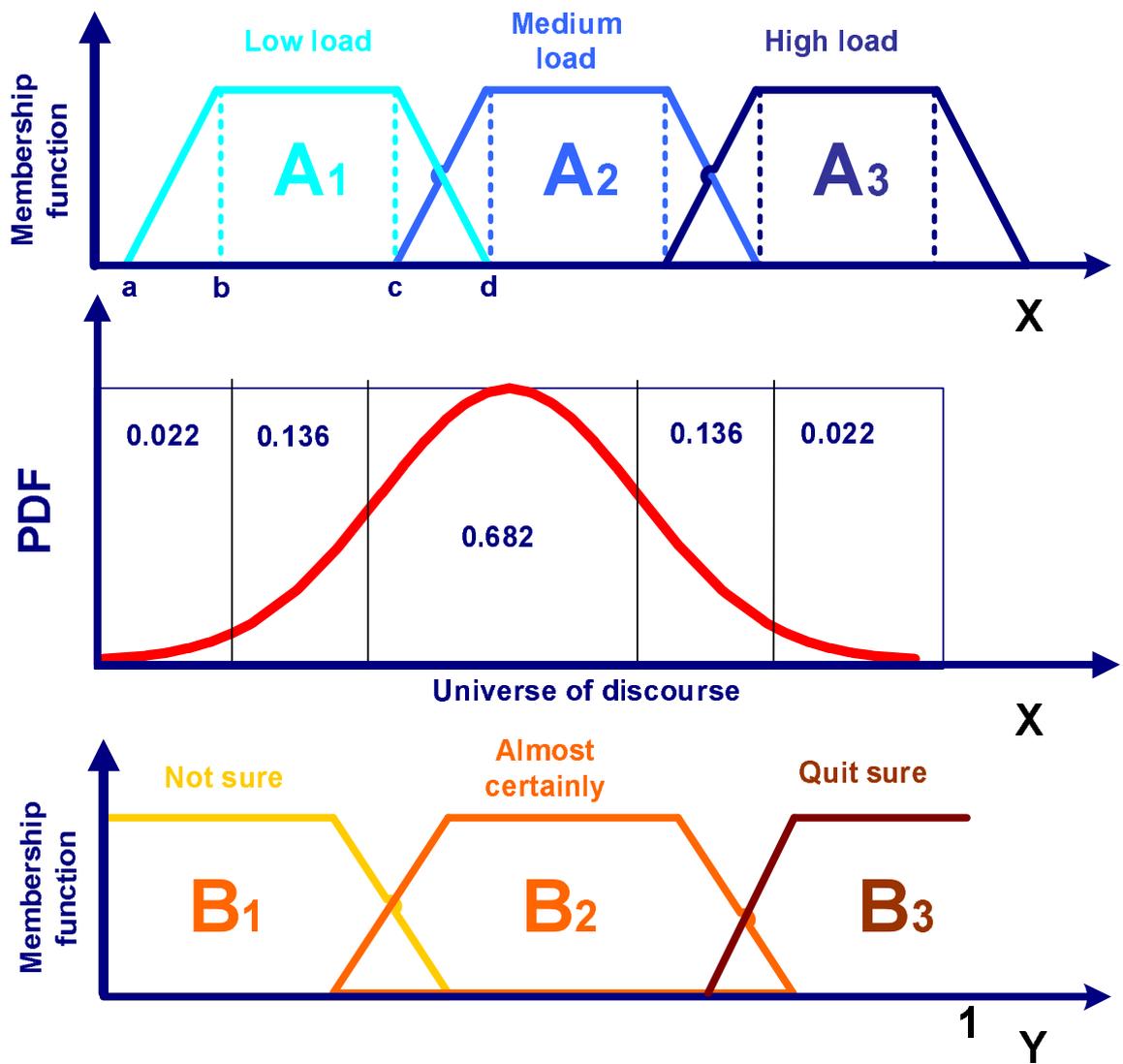}
	\caption{Concept of Z-number}
	\label{fig:shekl2}
\end{figure}

\end{document}